\documentclass{llncs}

\usepackage{graphicx}
\usepackage{multirow}

\begin{document}

\title{IMAGE SEGMENTATION AND CLASSIFICATION FOR SICKLE CELL DISEASE
USING DEFORMABLE U-NET}

\author{Mo Zhang\inst{1*}, Xiang Li\inst{2*}, Mengjia Xu\inst{3*}, Quanzheng Li\inst{2}}

\institute{$^{1}$Peking University, Beijing, China;\\
$^{2}$Harvard Medical School and Massachusetts General Hospital, Boston, MA;\\
$^{3}$Northeastern University, Shenyang, China. *Joint first authors}

\maketitle              

\begin{abstract}
Reliable cell segmentation and classification from biomedical images is a crucial step for both scientific research and clinical practice. A major challenge for more robust segmentation and classification methods is the large variations in the size, shape and viewpoint of the cells, combining with the low image quality caused by noise and artifacts. To address this issue, in this work we propose a learning-based, simultaneous cell segmentation and classification method based on the deep U-Net structure with deformable convolution layers. The U-Net architecture for deep learning has been shown to offer a precise localization for image semantic segmentation. Moreover, deformable convolution layer enables the free form deformation of the feature learning process, thus makes the whole network more robust to various cell morphologies and image settings. The proposed method is tested on microscopic red blood cell images from patients with sickle cell disease. The results show that U-Net with deformable convolution achieves the highest accuracy for segmentation and classification, comparing with original U-Net structure. 
\end{abstract}

\section{Introduction}
Sickle cell disease (SCD) is an inherited blood disorder, where patients with SCD have abnormal hemoglobin that can cause normal disc-shaped red blood cells (RBCs) to distort and generate heterogeneous shapes. The differences in cell morphology between healthy and pathological cells make it possible to perform the automatic cell segmentation and classification using image processing techniques, which is very important for faster and more accurate diagnosis of SCD. Various methods have been developed to perform RBC segmentation and/or classification, such as thresholding, region growing \cite{1}, watershed transform \cite{2}, deformable models \cite{3}, and clustering \cite{4}. However, traditional image processing models such as thresholding and region growing are susceptible to the noisy image background and blurred cell boundaries, which are common in microscopy images. Moreover, deformable models like active contour \cite{3} needs good initialization and relies on relatively clear cell morphology. In addition, due to the heterogeneous shapes and overlapped RBCs in SCD, recent open source cell detection tools, such as CellProfiler \cite{5}, CellTrack \cite{6} or Fiji \cite{7} are not readily to be used to accurately detect and classify the SCD RBCs. Hence, an effective SCD cell segmentation and classification method is still an open problem for the field.

Recently, deep learning methods with convolutional neural networks (CNN) have achieved remarkable success in the field of both natural image \cite{8} and medical image analysis \cite{9}. Among these methods, the fully convolutional network (FCN) has shown state-of-the-art performance in various real-world applications \cite{10}. Specifically, FCN has been applied in the cell segmentation problems \cite{11}\cite{12} and obtained good results. U-Net was developed based on FCN and takes skip connection between encoder and decoder into consideration, which has also been applied on medical images \cite{13}. Thus, towards our goal of simultaneous cell segmentation and classification (i.e. semantic segmentation), we envision that U-Net is a fitting solution as it can encode the variations in cell shape and texture, then perform the prediction (labeling) of the cells correspondingly. 

One of the major challenges in capturing the most discriminative shape and texture features of the RBCs is that cells can be imaged in various poses and sizes, thus a spatial-invariant scheme is needed to overcome those variations. For example, the work applies dense transformer network based on thin-plate spline, and has achieved superior performance on brain electron microscopy image segmentation problems \cite{14}. In this work, we apply deformable convolution \cite{15} to the U-Net architecture and develop the deformable U-Net framework for semantic cell segmentation. Deformable convolution accommodates geometric variations in the images by learning and applying adaptive receptive fields driven by data \cite{15}, in contrast to standard CNNs where the receptive field is constant. Therefore, it can be more robust to the spatial variations of the RBCs.

By training and testing the proposed framework on a manually-annotated microscopic imaging dataset consisting of both healthy and pathological cells, we perform the simultaneous segmentation and classification of the RBC in various experimental settings. The supreme accuracy for both segmentation and classification indicates that the proposed framework is a suitable solution for the automatic detection of SCD RBCs. To the best of our knowledge, this work is the first attempt of solving the SCD detection problem in an end-to-end semantic segmentation approach. 

\section{Method}

\subsection{Data acquisition and preprocessing}
The experimental blood sample is collected from UPMC (University of Pittsburgh Medical Center). 32 raw microscopy images are obtained using a Zeiss inverted Axiovert 200 microscope under $63\times$ oil objective lens using an industrial camera (Sony Exmor CMOS color sensor, 1080p resolution). Each microscopy image is in 4 color channels and the full image size is $1920\times1080$. More details about the acquisition protocol and processing can be found in \cite{16}. Due to the limitation of the GPU memory to load the entire image in, we divide each image into four parts and resize each part into a $256\times256$ square sample, to the total of 128 samples. No further image enhancement such as denoising and spatial filtering are used in this work.

\subsection{Deformable U-Net}

\begin{figure}
\centering
\includegraphics{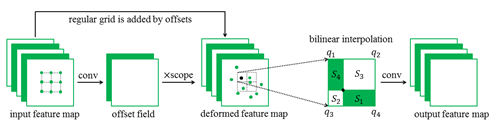}
\caption{\label{fig:1} Illustration of the deformable convolution, showing how the constant-sized feature maps are adaptively deformed.}
\end{figure}

We firstly implement a typical U-Net as the baseline model, which is an image-to-image classifier making pixel-wise predictions based on fully convolutional networks \cite{10}. Further, as standard convolution is inherently limited to deal with object shape transformations due to its regular square receptive field, we replace the convolution kernel with deformable convolution throughout the U-Net. Besides the technique of deformable convolution, there exist many methods to ensure the spatial invariance, such as data augmentation \cite{17} and spatial transformer networks (STN) \cite{18}. However, data augment is time consuming during the training as a much larger set of samples is needed to be generated and trained, while the global feature warping in STN cannot satisfy sophisticated vision tasks such as classification. On the other hand, deformable convolution can sample the input feature map in a local and dense way, and adaptive to the localization for objects with different shapes \cite{15}, which is exactly what we need in this work.

In the classic CNN architecture, convolution kernel is defined with fixed shape and size by sampling the input feature map on a regular grid. For example, the grid R for a $3\times3$ kernel is $R=\{(-1,-1),(-1,0),\cdots,(0,1),(1,1)\}$,For each pixel $p_0$ on the output feature map $y$ from image $x$, the standard convolution can be expressed as: 
\begin{equation}
y(p_0)=\sum_{p_n\in R}w(p_n)\cdot x(p_0+p_n),
\end{equation}
where $y(p_0)$ denotes the value on pixel $p_0$ in the output feature map, and $x(p_0+p_n)$ denotes the value on pixel $p_0+p_n$ in the input image.In contrast, deformable convolution adds 2D offsets to the regular sampling grid $R$, thus Eq.$(1)$ becomes:
\begin{equation}
y(p_0)=\sum_{p_n\in R}w(p_n)\cdot x(p_0+p_n+\Delta p_n).
\end{equation}
As offset $\Delta p_n$ is fractional probably, Eq.$(2)$ is implemented by bilinear interpolation as:
\begin{equation}
x(p)=\sum_{q}f(q_x,p_x)\cdot f(q_y,p_y)\cdot x(q), 
\end{equation}
where $p$ enumerates an arbitrary fractional location while $q$ denotes all integral locations on the input feature map. The one-dimensional kernel $f$ is defined as:
\begin{equation}
f(m,n)=max(0,1-|m-n|).
\end{equation}
Eq.$(3)$ is easy to compute as it is only related with the four nearest integral coordinates $q_i,i=[1,2,3,4]$ of $p$. Eq.$(3)$ is also equivalent to:
\begin{equation}
x(p)=\sum_{i=1}^{n}x(q_i)\cdot S_i,
\end{equation}
where $S_i,i=[1,2,3,4]$ is the area of the assigned rectangle generated by $q_i,i=[1,2,3,4]$ and $p$. 

The detailed procedure of deformable convolution is described in Fig. 1. Firstly, we implement an additional convolution with activation function TANH to learn offset field from the input feature map, which are normalized to $[-1,1]$. The offset field has the same height and width with input feature map while its number of channels is $2N (N=|R|)$. The offset field is then multiplied by parameter $s$ (which is used to adjust the scope of receptive field) and added by the regular grid R to obtain the sampling locations (every coordinate on offset field has $N$ pairs of values corresponding to the regular grid $R$). Finally, values of the irregular sampling coordinates are computed via bilinear interpolation, then the original convolution kernel samples the deformed feature map to get the new feature map. In this work, we set deformable kernel works in the same way across different channels, rather than learning a separate kernel for each channel to improve the learning efficiency. The deformable U-Net can be easily trained end-to-end (from the input image to the label map) through back propagation in the same way with the U-Net architecture. 

\subsection{Network structure and implementations}

The architecture of the deformable U-Net is shown in Fig. 2. It consists of an encoder path (left side) and a decoder path (right side) each with three layers. In the encoder path, each layer has two $3\times3$ deformable convolutions followed by a $2\times2$ max pooling operation with stride of 2, which doubles the number of channels and halves the resolution of input feature map for down-sampling. The encoder is followed by two $3\times3$ deformable convolutions called bottom layers. Each step in the decoder path contains a $3\times3$ deconvolution with stride 2 followed by two $3\times3$ deformable convolutions corresponding to its counterpart in the encoder except for the last layers where the label map is predicted. For the task of segmentation, the dimension of the last layer is $256\times256\times2$. For the classification with two cell types, the dimension should be $256\times256\times3$. The skip connection between encoder and decoder helps to preserve more contextual information for better localization \cite{13}. Structure for the baseline classic U-Net is the same as in Fig. 2, only with the deformable convolution kernels replaced by the standard convolution kernels. We use RELU as activation function with scope of 2 and batch normalization for convolution operations. The network is optimized by the Adam algorithm with initial learning rate $10^{-3}$ and weight decay $10^{-8}$. For both deformable and baseline U-Net training, we use the same batch size (5) and epoch (10000). The network is implemented in TensorFlow 1.2.1.

\begin{figure}
\centering
\includegraphics{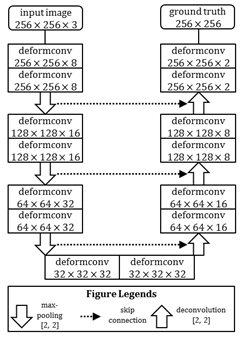}
\caption{\label{fig:2} Architecture of the deformable U-Net in this work.}
\end{figure}

\section{Results}

\subsection{Experimental evaluation of the model performance}

\begin{table}
\begin{center}
\caption{Model performance comparison for the task of cell segmentation and classification.}
\begin{tabular}{|c|c|c|c|c|c|c|c|c|}
\hline
\ &\multicolumn{3}{|c|}{Segmentation}&\multicolumn{5}{|c|}{Classification} \\
\ &\multicolumn{3}{|c|}{(totally 490 cells)}&\multicolumn{5}{|c|}{(totally 510 cells)} \\
\hline
Model  & Loss & FN & FP & Loss & FN & FP & Error\uppercase\expandafter{\romannumeral1} & Error\uppercase\expandafter{\romannumeral2} \\
\hline
U-Net  & 0.0545 & 9 & 17 & 0.1537 & 6 & 15 & 71 & 45 \\
\hline
Deformable U-Net  & \textbf{0.0509} & \textbf{5} & \textbf{6} & \textbf{0.1012} & \textbf{3} & \textbf{5} & \textbf{68} & \textbf{12} \\
\hline
\end{tabular}
\end{center}
\end{table}

To evaluate the performance of the deformable U-Net as well as the baseline U-Net model, we perform the cell segmentation and classification task on the SCD microscopic imaging dataset as introduced in 2.1. Model accuracy is mainly evaluated on the cell-level testing results
by manually comparing the output image from U-Net with the ground truth. For cell segmentation, we use measurements inducing “loss” which indicates pixel-level accuracy, “false negative (FN) rate” counting the number of under-discovered cells and “false positive (FP) rate” counting the number of falsely-labeled cell to summarize the results. For cell classification, in addition to the above measurements, we calculate the Error I rate which indicates the number of misclassified cells (healthy as pathological or vice-versa). Further, as certain cells are segmented out yet identified as two classes simultaneously (i.e. at least one quarter of the total area of cell is labeled as one class while the rest of the area is labeled as the other class), we calculate the Error II rate to measure such kind of errors. In this work, we perform the experiment for segmentation and classification on two separate datasets randomly sample from the whole dataset (128 samples from 32 images).Both experiments use 88 random samples for training and the rest 40 samples for testing. The quantitative evaluation is shown in Table 1. Bold value indicates the better performance between the two models.

As indicated in the numbers, both networks can achieve good cell-level segmentation accuracy (97.8\% for deformable U-Net, 94.7\% for U-Net) and reasonable classification accuracy (82.7\% for deformable U-Net, 73.1\% for U-Net), while deformable U-Net outperforms the baseline model for all the evaluation criteria. Specifically, deformable U-Net has much less false positives for the segmentation task, indicating that it is more robust to the background noise presented in the microscopic images. One example can be found in the top row of Fig. 3, where the baseline U-Net mislabels a background object as the cell while deformable U-Net predict the accurate negative label.

For the cell classification task, as visualized in Figure 4, cells colored in red are pathological cells (usually sickle-shaped and/or with different texture) while cell colored in green are normal (usually disc-shaped). Both models can capture such differences in the shape and texture feature of the cells to make reasonable classification. Unlike methods in the earlier literatures such as \cite{16}, both the deformable and baseline U-Net are trained and applied end-to-end without any explicit feature extractions for the purpose of classification, indicating that the discriminative features are automatically learned within the training. Further, the deformable U-Net shows better performance than the baseline model as it can maintain the relatively stable classification label for the whole cell, thus achieve a one-third Error II rate (12) comparing with the U-Net (45).

\begin{figure}
\centering
\includegraphics{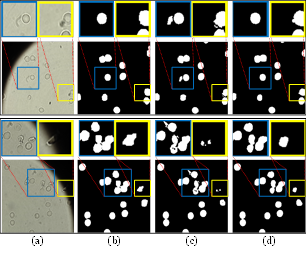}
\caption{\label{fig:3} Cell segmentation results from two sample images shown in the top and bottom panels. The top row in each panel shows the corresponding zoom-in view. The four columns in the figure are: (a) Raw image; (b) Ground truth; (c) Prediction of U-Net; (d) Prediction of deformable U-Net.}
\end{figure}

\subsection{Discussion on the effect of deformable kernel}

Based on the testing results as summarized in 3.1, we would like to discuss the possible causes for the differences in the performance of deformable and baseline U-Net. Firstly, from the prediction maps we have found that the False Positive (FP) cells segmented by the baseline U-Net are usually caused by the fact that the network recognizes the bright noise in the background as RBCs. While deformable U-Net can avoid such mistakes because the noise objects usually have smaller size than the true cells. From the network structure perspective, the only difference between the two models lies in the extra deformable kernel, it is possible that deformable kernel can help learning more spatial features to help it capturing the size information of object. Secondly, it can be observed that deformable U-Net produces a smoother cell boundary comparing with baseline U-Net, indicating it has better capability of dealing with the object edges, mainly thanks to the more flexible receptive fields provided by the deformation operation. Such capability also leads to the better Error II rate of the deformable U-Net, as it recognizes more accurate boundary of the prediction to form an integrated cell labeling without any shape prior. 

\begin{figure}
\centering
\includegraphics{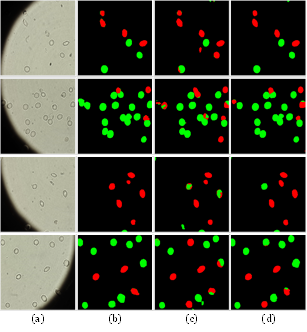}
\caption{\label{fig:4} Cell classification results from four sample images shown in each row. (a) Raw image; (b) Ground truth; (c) Prediction of U-Net; (d) Prediction of deform U-Net.}
\end{figure}

\section{Discussion}

In this work, we propose a deep learning framework to simultaneously segment and classify red blood cells for the sickle cell disease. Experimental results show that the deformable U-Net structure used in the proposed framework has superior performance than the state of art semantic segmentation U-Net algorithm. The framework is more robust to the different variations in cell size, texture and shape, which is reflected in its ability in discriminating the noise object and high consistency in performing the prediction on cell boundaries. The computation time for training the deformable U-Net is almost four times more than the baseline U-Net, the time cost is acceptable and the predicting speed which is most important for application is almost the same for the two networks. In the next step, we will increase the size of the training dataset to support fine-grained classification with more cell types (i.e. more labels for the pathological cells), in order to build a robust automatic system for SCD detection and diagnosis, readily to be applied in the clinical practice.

\section{Acknowledgement}

This work is supported by the MGH \& BWH Center for Clinical Data Science. We gratefully acknowledge the help of algorithm development from the Data Integration, Visualization, and Exploration (DIVE) Laboratory in Washington State University.


%
%

\end{document}